# A Greedy Algorithm for Low-Crossing Partitions for General Set Systems

Mónika Csikós[1]     Alexandre Louvet[2]     Nabil Mustafa[2]

**Abstract**

Simplicial partitions are a fundamental structure in computational geometry, as they form the basis of optimal data structures for range searching and several related problems. Current algorithms are built on very specific spatial partitioning tools tailored for certain geometric cases. This severely limits their applicability to general set systems. In this work, we propose a simple greedy heuristic for constructing simplicial partitions of any set system. We present a thorough empirical evaluation of its behavior on a variety of geometric and non-geometric set systems, showing that it performs well on most instances. Implementation of these algorithms is available on Github (C++, Rust)[3].

## 1 Introduction

Let $(X, \mathcal{F})$ be a set system, where $X$ is a finite set of $n$ elements and $\mathcal{F}$ is a collection of $m$ subsets of $X$. We refer to the elements of $\mathcal{F}$ as *ranges*. We say that a range $F \in \mathcal{F}$ *crosses* a set $P \subseteq X$ if and only if there exist two elements $x, y \in P$ such that $x \in F$ and $y \notin F$. Let $I(P, F)$ be the indicator function that is 1 iff $F$ crosses $P$, and 0 otherwise. Given a set system $(X, \mathcal{F})$, our goal is to construct a partition $\mathcal{P}$ of $X$ such that each set is of approximately the same size, and each range in $\mathcal{F}$ crosses a sub-linear number of parts of $\mathcal{P}$. These two properties are formalized in the following definitions.

**Definition 1.1. (Crossing Number)** Given $(X, \mathcal{F})$ and a partition $\mathcal{P} = \{P_1, ..., P_t\}$ of $X$, the crossing number of $\mathcal{F}$ with respect to $\mathcal{P}$, denoted by $\kappa_\mathcal{F}(\mathcal{P})$, is the maximum number of sets of $\mathcal{P}$ that are crossed by a range in $\mathcal{F}$. Formally,

$$\kappa_\mathcal{F}(\mathcal{P}) = \max_{F \in \mathcal{F}} \sum_{i=1}^{t} I(P_i, F).$$

**Definition 1.2. (($\tau, \kappa$)-Partitions)** Given a set system $(X, \mathcal{F})$, $n = |X|$, a $(t, \kappa)$-partition of $(X, \mathcal{F})$ is a partition of $X$ into disjoint sets $P_1, ..., P_t$ such that

1. for all $i \in [t-1]$, we have $|P_i| = \lfloor \frac{n}{t} \rfloor$,
2. $\frac{n}{t} \leq |P_t| \leq \frac{2n}{t}$, and
3. $\kappa_\mathcal{F}(\mathcal{P}) \leq \kappa$.

We refer to the $P_i$'s as parts of the partition $\mathcal{P}$.

---

[1]Université Paris Cité, IRIF, CNRS UMR 8243. Supported by the program "Investissement d'Avenir" launched by the French Government and implemented by ANR, with the reference ANR-18-IdEx-0001 as part of its program Emergence for the project VC-GRAPHES.

[2]Université Sorbonne Paris Nord, CNRS, Laboratoire d'Informatique de Paris Nord, LIPN, F-93430 Villetaneuse, France. Supported by the French ANR PRC grant ADDS (ANR-19-CE48-0005).

[3]The difference between the two implementations is detailed in Section 5.1

**Previous work.** The study of $(t, \kappa)$-partitions originated in computational geometry in the late 1980s, under the name *simplicial partitions*. A breakthrough result that established a key bound, as well as their significance, is the following.

**Theorem 1.3. [Mat92]** Given a set $X$ of $n$ elements in $\mathbb{R}^d$, let $\mathcal{H}$ denote the set system induced on $X$ by the family of half-spaces in $\mathbb{R}^d$. Then for any integer parameter $t \in [n]$, there exists a $\left(\Theta(t), O\left(t^{1-\frac{1}{d}}\right)\right)$-partition of $(X, \mathcal{H})$.

**Remark.** The original proof of [Mat92] actually implies that $X$ is partitioned into $t-1$ parts of size $\lfloor \frac{n}{t} \rfloor$ and one part of size between $\frac{n}{t}$ and $\frac{2n}{t}$.

On the other hand, not all set systems—even very simple geometric ones—admit partitions for all parameters $t$. This was shown in 1987 by Alon, Haussler and Welzl [AHW87], for the case of the set system induced by lines in the projective plane[4].

Simplicial partitions are a key tool for the geometric range searching problem, where the aim is to preprocess a set of points $X$ (in $\mathbb{R}^d$) to be able to answer simplex containment queries on $X$. That is, each query specifies a simplex $Q \subseteq \mathbb{R}^d$, and the goal is to compute the set $Q \cap X$ quickly. The current-best way to solve this problem is via hierarchical data-structures (e.g., partition trees), which are derived from recursive applications of simplicial partitions [Mat92, Mat93]. This construction was further improved and simplified by Chan [Cha12].

Most previous algorithms for computing partitions with sub-linear crossing numbers rely on two main ingredients: the multiplicative weights update method, and the following key spatial partitioning tool, called cuttings.

**Definition 1.4. (Cuttings)** Given a set $\mathcal{H}$ of $m$ hyperplanes in $\mathbb{R}^d$ and a parameter $r > 0$, a $\frac{1}{r}$-cutting for $\mathcal{H}$ is a collection of (possibly unbounded) $d$-dimensional closed simplices with disjoint interiors, which together cover $\mathbb{R}^d$ and such that the interior of each simplex intersects at most $\frac{m}{r}$ hyperplanes of $\mathcal{H}$.

The best known algorithm to compute a $\frac{1}{r}$-cutting of size $O(r^d)$ has a running time of $O(mr^{d-1})$ [Cha93]. Moreover, a robust implementation of the algorithm is non-trivial, currently only available in $\mathbb{R}^2$ [Har00].

On the experimental side, for the purposes of computing $\varepsilon$-approximations, Matheny and Phillips [MP18] constructed simplicial partitions for the specific case of set systems induced by half-spaces in $\mathbb{R}^2$, via methods inspired by the algorithms of Matoušek [Mat92] and Chan [Cha12]. However these methods rely on cuttings; therefore all previous experimental studies were limited to half-spaces in $\mathbb{R}^2$.

**Our results.** In this paper, we consider the problem of computing partitions with low crossing numbers for general set systems. We aim for a simple algorithm that is fast in practice. In particular, our method does not rely on cuttings and so it works for general set systems, as well as geometric set systems in higher dimensions.

Our algorithm will construct a low-crossing partition iteratively, building one part at a time by greedily extending it with one element which does not introduce too many new crossings. As a motivation for this greedy approach, we prove that (under some hereditary assumptions), any $\left(t, O\left(t^{1-\frac{1}{d}}\right)\right)$ partition can be created using this method. In order to ensure low crossing number globally, we apply multiplicative weight updates (MWU) between parts. The resulting algorithm has time complexity $O(nmt)$. We also give a faster variant of this algorithm with running time $\tilde{O}(mn + nt^2)$ that can, further, be partially parallelized on as many as $m$ cores.

---
[4]The precise definition and some experiments on this set system are presented in the Appendix.

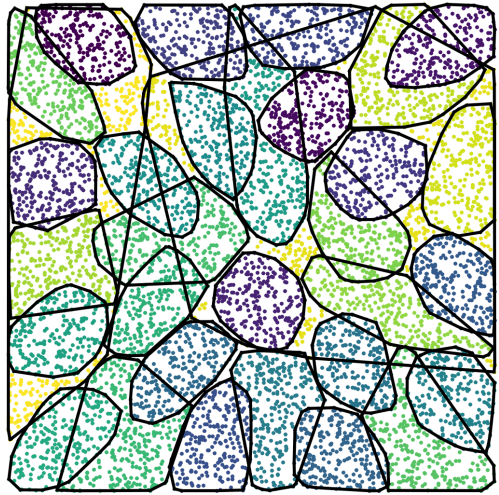

We evaluate our algorithm for a variety of set systems, including abstract and high-dimensional geometric ones. As an illustration, the figure above shows a partition, of size 32, constructed by our algorithm (MinWeight) for the set system induced by half-spaces in $\mathbb{R}^2$. All the elements of the same part are drawn the same color, the first constructed part's elements are in purple and the elements of the last one in yellow. The black lines mark the convex hull of each part.

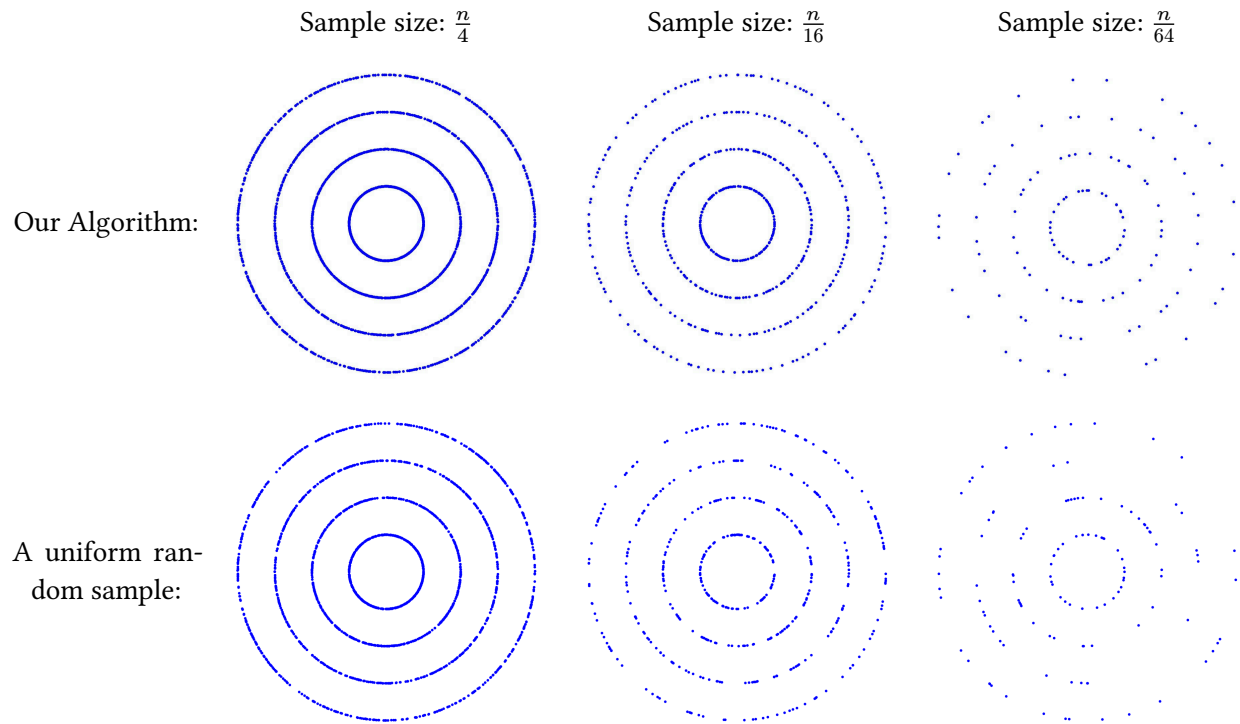

Figure 1: Approximations of a set system with 8192 elements on concentric circles and ranges induced by disks. On top, partitions computed with MinWeight (with 1024 random disks) and on the bottom with a uniform random sample.

**Applications.** A classical use of simplicial partitions is for combinatorial data approximation, in particular, for computing sub-quadratic sized $\varepsilon$-approximations (see Section 5.3.4 for the definition and more details). As an illustration, we compare the results of a uniform random $\varepsilon$-approximation to an $\varepsilon$-approximation computed with our algorithm, for the geometric set system induced by disks. We see on Figure 1 that the approximation we obtain with a partition is visibly better, leaving less gaps on the concentric circles, than a uniform random sample of the same size. We present a more detailed comparison of these methods in Section 5.3.4. Further applications and their thorough evaluations will be presented in the full version of the paper.

**Organization of the paper.** In Section 2, we prove a theorem that motivates the choice of a greedy approach to the problem. In Section 3 and Section 4, we detail our algorithms and in Section 5, we provide data on the effectiveness of our algorithms on different types of set systems (both abstract and geometric).

## 2 The Ordering Theorem

Let $(X, \mathcal{F})$ be a set system for which there exists a partition $\mathcal{P}$ of size $t$ that has a low crossing number with respect to $\mathcal{F}$. Our main insight is that, for each $P_i \in \mathcal{P}$, there exists a permutation of the elements of $P_i$ that can be added in sequence, *iteratively*, such that the crossing number increase for each addition is upper-bounded by a specific function. Thus, following such a sequence of additions results in a set of $\frac{n}{t}$ points that is crossed by few ranges. We will refer to this function, derived below, as the *potential function*. The only requirement that we need for the existence of such a good permutation is the following hereditary property: there is a constant $d \geq 1$ such that

$$\forall Y \subseteq X \text{ and all } s \in [|Y|], (Y, \mathcal{F}|_Y) \text{ admits an } \left(s, s^{1-\frac{1}{d}}\right)\text{-partition} \tag{1}$$

where $\mathcal{F}|_Y = \{F \cap Y : F \in \mathcal{F}\}$.

Note that Equation 1 is satisfied for those geometric set systems where partitions of sub-linear crossing numbers are proven to exist (e.g., geometric set systems induced by semialgebraic sets [AMS13]).

**Theorem 2.1.** Let $(X, \mathcal{F})$ be a set system satisfying Equation 1 and $\mathcal{P} = \{P_1, ..., P_r\}$ be $r$ disjoint subsets of $X$, where $|P_i| = \frac{n}{t}$ for all $i \in [r]$. Let $\mathcal{R}$ be a family of subsets of $X$ with crossing number at most $\kappa$ with respect to $\mathcal{P}$. Let $P_l \in \mathcal{P}$ be any part and let $\mathcal{R}_l$ denote the set of ranges crossing $P_l$. Then there exists an ordering of the elements of $P_l$, say $\langle x_1, x_2, ..., x_{\frac{n}{t}} \rangle$, such that:

$$\forall k \in \left[\frac{n}{t}\right], \text{ the prefix set } \{x_1, ..., x_k\} \text{ is crossed by at most } \frac{4|\mathcal{R}_l|t^{1/d}k^{1/d}}{n^{1/d}} \text{ sets of } \mathcal{R}. \tag{2}$$

Moreover, if $P_l$ is chosen uniformly at random from $\{P_1, ..., P_r\}$, then with probability at least $\frac{1}{2}$, there exists an ordering $\langle x_1, x_2, ..., x_{\frac{n}{t}} \rangle$ of the elements of $P_l$ such that

$$\forall k \in \left[\frac{n}{t}\right], \text{ the prefix set } \{x_1, ..., x_k\} \text{ is crossed by at most } \frac{4|\mathcal{R}|\kappa t^{1/d}k^{1/d}}{rn^{1/d}} \text{ sets of } \mathcal{R}. \tag{3}$$

We will use Theorem 2.1 to define a suitable potential function: each partition will be constructed greedily by adding elements to it that satisfy the upper bound in Equation 3. Towards this goal, in Theorem 2.1, we can re-try with several random starting elements, in case the algorithm fails to compute an ordering satisfying the bound of Equation 3, see Section 3 for more details.

We remark that the ordering provided by Theorem 2.1 is not trivial, even in simple set systems. For instance consider the set system on the $n$ elements of the integer grid $[0, \sqrt{n}] \times [0, \sqrt{n}]$ in $\mathbb{R}^2$ and ranges defined by half-planes bounded by $\sqrt{n} - 1$ evenly-spaced horizontal lines and $\sqrt{n} - 1$ evenly-spaced vertical lines over the grid. We consider two orderings starting from the origin, in the construction of a single partition; see Figure 2. On the left, the first $k$ elements of the ordering are always contained within a box of side-length) $\sqrt{k}$. Thus, the number of ranges crossed by the first $k$ elements is proportional to $\sqrt{k}$, as desired. This demonstrates that the origin is a good starting element of the ordering. On the other hand, the right-side drawing demonstrates a *bad* ordering where the number of lines crossed increases linearly with the number of elements in the prefix of the ordering.

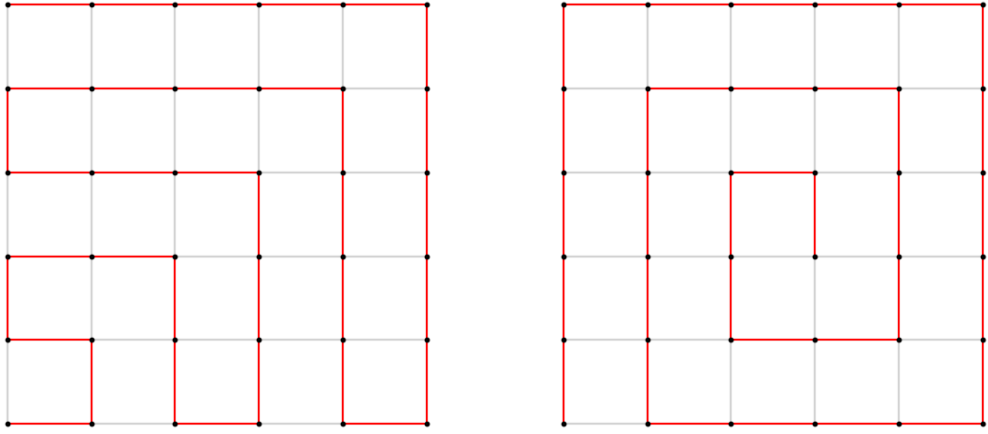

Figure 2: On the left, a *good* ordering with a number of crossings proportional to $\sqrt{k}$. On the right, a *bad* ordering where the crossing number evolves linearly with $k$, for $1 \leq k \leq 2\sqrt{k}$.

We return to the proof of Theorem 2.1.

*Proof.* Set $Q^0 = P_l$ and $\mathcal{S}^0 = \{R \cap Q^0 : R \in \mathcal{R}_l\}$. By applying Equation 1 with $s = 2^d$, $(Q^0, \mathcal{S}^0)$ has a simplicial partition $\mathcal{P}^1$, of size $2^d$, with crossing number at most

$$\left(2^d\right)^{1-\frac{1}{d}} = 2^{d-1}.$$

By the pigeonhole principle, there exists a part, say $Q^1 \in \mathcal{P}^1$, that is crossed by at most

$$\frac{|\mathcal{S}^0| \cdot 2^{d-1}}{2^d} = \frac{|\mathcal{S}^0|}{2} \leq \frac{|R_l|}{2}$$

ranges of $\mathcal{R}$. Denote the set of the ranges that cross $Q^1$ by $\mathcal{S}^1$. Note that $\frac{n}{2^d t} \leq |Q^1| \leq \frac{n}{2^{d-1} t}$. Now we repeat the same process with $(Q^1, \mathcal{S}^1)$. That is, at the $j$-th step, we compute a $(2^d, 2^{d-1})$-partition, denoted by $\mathcal{P}^{j+1}$, of $(Q^j, \mathcal{S}^j)$. Then by pigeonhole principle, there exists a set $Q^{j+1} \in \mathcal{P}^{j+1}$ that is crossed by at most

$$\frac{|S^j| \cdot 2^{d-1}}{2^d} = \frac{|S^j|}{2} \leq \frac{|R_l|}{2^{j+1}}$$

ranges of $\mathcal{R}$. Further, we have $\frac{n}{2^{(j+1)d}t} \leq |Q^{j+1}| \leq \frac{n}{2^{(j+1)(d-1)}t}$. We continue as long as

$$|Q_j| < 2^d$$

and denote $T$, the first index where this is true. This results in a sequence
$$Q^T \subseteq Q^{T-1} \subseteq ... \subseteq Q^0 = P_l.$$

Our final ordering, denoted by $\pi$, is as follows:

- elements of $Q^T$ (in any order),
- the elements of $Q^{T-1} \setminus Q^T$ (in any order),
- then the elements of $Q^{T-2} \setminus Q^{T-1}$ (in any order),

  $\vdots$

- finally, the elements of $Q^0 \setminus Q^1$.

Now fix any $k \in \left[\frac{n}{t}\right]$, and let $j \in [T]$ be the largest index such that $|Q^j| > k$. That is,

$$\frac{n}{2^{(j+1)d}t} \leq |Q_{j+1}| \leq k < |Q_j| \leq \frac{n}{2^{j(d-1)}t}. \tag{4}$$

The first $k$ elements in our ordering $\pi$ all lie in $Q^j$, and by our construction, $Q^j$ is crossed by at most $\frac{|R_l|}{2^j}$ sets of $\mathcal{R}$. That is, the set formed by the $k$ first elements in our ordering is crossed by at most these many sets of $\mathcal{R}$:

$$\begin{aligned}
\frac{|R_l|}{2^j} &= \frac{2|R_l|}{\left(2^{(j+1)d}\right)^{1/d}} \frac{t^{1/d}n^{1/d}}{t^{1/d}n^{1/d}} \\
&= \frac{2|R_l|t^{1/d}}{n^{1/d}} \left(\frac{n}{2^{(j+1)d}t}\right)^{1/d} \\
&\leq \frac{2|R_l|t^{1/d}}{n^{1/d}} \cdot k^{1/d},
\end{aligned}$$

where the last step follows from Equation 4.

Finally if $P_l$ is a part picked uniformly at random, we have

$$\begin{aligned}
\mathbb{E}[|\mathcal{R}_l|] &= \sum_{i=1}^r \frac{1}{r} |\mathcal{R}_i| \\
&= \frac{1}{r} \sum_{R \in \mathcal{R}} \sum_{i=1}^r I(P_i, R) \\
&\leq \frac{1}{r} \sum_{R \in \mathcal{R}} \kappa \\
&= \frac{|\mathcal{R}|\kappa}{r}.
\end{aligned}$$

By Markov's inequality, we have

$$\Pr\left[|\mathcal{R}_l| \leq \frac{2|\mathcal{R}|\kappa}{r}\right] \leq \frac{1}{2}, \tag{5}$$

Thus with probability at least $\frac{1}{2}$, the $k$ first elements in our ordering are crossed by at most

$$\frac{2|\mathcal{R}_l|t^{\frac{1}{d}}}{n^{\frac{1}{d}}} \cdot k^{\frac{1}{d}} \leq \frac{4|\mathcal{R}|\kappa t^{\frac{1}{d}} k^{\frac{1}{d}}}{rn^{\frac{1}{d}}}$$

ranges of $R$. □

# 3 Our Greedy Algorithm Using the Potential Function

Classical methods for building simplicial partitions use the multiplicative weight update (MWU) framework to maintain a weight function on each $F \in \mathcal{F}$ that evolves with the number of parts crossed by $F$. This is combined with the key step of finding a good set of $\frac{n}{t}$ elements of $X$ (which constitutes the next part) that is crossed by ranges of low total weight in each iteration.

**Greedy Potential.** In our method, we keep the MWU framework to ensure low crossing number, but take a different approach for constructing the parts, inspired by Theorem 2.1. At iteration $i$, we sample a random element $x_0 \in X \setminus (P_1 \cup ... \cup P_{i-1})$ to be the starting element of $P_i$. The algorithm proceeds by greedily adding elements to $P_i$ so that the total weight of ranges crossing $P_i$ stays below the potential function bound of Theorem 2.1. To this end, we maintain a function $\omega(\cdot)$ which stores, for each $x \in X \setminus (P_1 \cup ... \cup P_i)$, the cost of adding $x$ to $P_i$. In other words, $\omega(x)$ is the total weight of ranges in

$$\mathcal{C}(P_i, x) := \text{ranges in } \mathcal{F} \text{ that do not cross } \mathcal{P}_i \text{ but cross } P_i \cup \{x\}.$$

Initially, for any $x$, $\omega(x)$ is equal to the total weight of ranges crossing the edge $\{x_0, x\}$. Note that each time we pick an element $x'$ to be added to $P_i$, we need to adjust, for each $x \in X \setminus P_i$, its weight $\omega(x)$ by removing the weight of those ranges that are both in $\mathcal{C}(P_i, x)$ and $\mathcal{C}(P_i, x')$.

Formally, at step $k$ of the construction of $P_i$, with $x_0, ..., x_{k-1}$ the elements of $X$ selected in the first $k-1$ steps of the construction of $P_i$ and $\pi(F) = 2^{\sum_{j=1}^{i-1} I(P_j, F)}$.

$$\omega(x) = \sum_{F \in \mathcal{F}} \pi(F) I(\{x_0, x\}, F)(1 - I(\{x_0, ..., x_{k-1}\}, F)),$$

The resulting algorithm is presented in GreedyPotential. When set systems admit partitions with sublinear crossing number, $d$ is immediately deduced from the crossing number. Otherwise, it is possible to run the algorithm $\log(n)$ times to search the value of $d$ in $[1, n]$ that gives the best crossing number.

Interestingly, this algorithmic idea is already present in Chan's paper [Cha12], though it is used *between* partitions. Chan's algorithm starts from $X$ and then iteratively *refines* the initial partition with the use of cuttings. The order in which the partitions are refined is via a random permutation. Chan's algorithm is top-down, requiring cuttings to do the refinement from one level to the next. The heuristic we propose constructs the parts bottom-up via the existence of a potential function that guides our greedy algorithm.

The classical proofs proceed by upper-bounding, using cuttings or packing lemmas, the number of ranges crossing the $i$-th constructed part. This upper-bound is an absolute bound depending only on $n, m, d$, and $i$. For general set systems, we do not have access to cuttings or packing lemmas, and we only rely on the fact that the input set system satisfies Equation 1. Thus we take a different strategy in the proof: we derive the upper bound on the crossing number for the $i$-th part by applying Equation 1 iteratively to the remaining elements. Note that this upper bound depends also on the crossing number of the parts constructed so far. However, the theorem below shows that this additional term adds only a logarithmic factor to the crossing number.

**Algorithm 1:** GreedyPotential

```
1  n ← |X|, m ← |F|, P ← ∅
2  ∀F ∈ F, π(F) ← 1
3  for i ← 1 to t do
4      x_0 ← a random element of X
5      P_i ← {x_0}
6      cost ← 0
7      for F ∈ F do
8          foreach x ∈ X with F crossing {x_0, x} do
9              ω(x) ← ω(x) + π(F)
10     for k ← 2 to n/t do
11         y_k ← any element of X s.t. cost + ω(y_k) ≤ (2k^{1/d} Σ_{F∈F} π(F)) / |X|^{1/d}
12         X ← X \ {y_k}
13         cost ← cost + ω(y_k)
14         foreach F ∈ C(P_i, y_k) do
15             foreach x ∈ X with F crossing {x_0, x} do
16                 ω(x) ← ω(x) − π(F)
17         P_i ← P_i ∪ {y_k}
18     ∀F ∈ F, π(F) ← π(F) · 2^{I(P_i,F)}
19     P ← P ∪ {P_i}
20     X ← X \ P_i
21 return P
```

**Theorem 3.1. Overall crossing number bound.** Let $(X, \mathcal{F})$ be a set system satisfying Equation 1. Assuming GreedyPotential is always able to pick an element satisfying Equation 3, then GreedyPotential constructs a $\left(t, O\left(\ln m + t^{1-\frac{1}{d}} \ln t\right)\right)$-partition w.r.t. $\mathcal{F}$.

*Proof.* Let $X'$ be a subset of $X$ of size exactly $t \lfloor \frac{n}{t} \rfloor$. Let $\mathcal{P}_0$ be a partition of $X'$ into $t$ equal-sized subsets, with crossing number $t^{1-\frac{1}{d}}$. Applying Theorem 2.1, we pick a random element from $X'$—which is equivalent to picking a random part of $\mathcal{P}_0$ as they all have the same size—and construct an $\frac{n}{t}$-sized set, say $S_1$, containing it using a greedy algorithm. Since Equation 3 is satisfied for $S_1$, the number of ranges of $\mathcal{F}$ crossing $S_1$ is at most

$$\frac{4|\mathcal{F}|\, t^{1-1/d} t^{1/d} (n/t)^{1/d}}{t n^{1/d}} = \frac{4|\mathcal{F}|}{t^{1/d}}.$$

Next, we construct a new family of multisets $\mathcal{F}_1$ by duplicating the ranges crossing $S_1$:

$$\mathcal{F}_1 = \mathcal{F} \cup \{F \in \mathcal{F} : F \text{ crosses } S_1\}.$$

By our assumption on $(X, \mathcal{F})$, there exists a partition $\mathcal{P}_1$ of size $t-1$ for $X' \setminus S_1$ with crossing number $(t-1)^{1-1/d}$. Apply Theorem 2.1 and the greedy algorithm to get a set $S_2$ of $\frac{n}{t}$ elements from $X' \setminus S_1$. Since Equation 3 is again satisfied for $S_2$, the number of ranges of $\mathcal{F}_1$ crossing $S_2$ is at most

$$\frac{4|\mathcal{F}_1|\,(t-1)^{1-1/d}t^{1/d}\left(\frac{n}{t}\right)^{1/d}}{(t-1)n^{1/d}}$$

$$\leq \frac{|\mathcal{F}|\left(1+\frac{4}{t^{1/d}}\right)\left(4t^{1-1/d}\right)}{t-1}.$$

We again duplicate the sets of $\mathcal{F}_1$ crossing $S_2$, to get the next multiset $\mathcal{F}_2$.

Continuing on for $t$ steps, we get a partition $\{S_1, ..., S_t\}$ of $X'$, and $|\mathcal{F}_t|$ can be bounded as

$$|\mathcal{F}_t| \leq |\mathcal{F}|\prod_{i=1}^{t}\left(1+\frac{4t^{1-1/d}}{i}\right)$$

$$\leq m\exp\left(4t^{1-1/d}\sum_{i=1}^{t}\frac{1}{i}\right)$$

$$= O\!\left(m\exp(t^{1-1/d}\ln t)\right).$$

On the other hand, a range crossing $l$ parts in $\{S_1, ..., S_t\}$ appears at most $2^l$ times in $\mathcal{F}_t$, implying that

$$l \leq \log(|\mathcal{F}_t|) = O\!\left(\ln m + t^{1-1/d}\ln t\right).$$

$\{S_1, ..., S_t \cup (X \setminus X')\}$ is a $\left(t, O(\ln m + t^{1-1/d}\ln t)\right)$-partition of $X$ w.r.t. $\mathcal{F}$ because its crossing number is at most one more than the crossing number of $\{S_1, ..., S_t\}$ (the set with maximum crossing number might intersect $S_t \cup (X \setminus X')$ but not $S_t$). □

Note that the first set $S_1$ may contain elements from any mixture of the sets of $\mathcal{P}_0$. This is not a problem: the only property that we require is an upper bound on the total number of sets of $\mathcal{F}$ that cross $S_1$. The next step, for computing $S_2$, can take as input an arbitrary partition $\mathcal{P}_1$ on $X \setminus S_1$ with a suitably low crossing number. We do not require any "consistency" between the partitions $\mathcal{P}_0$ and $\mathcal{P}_1$.

The reader may notice that in our experiments we use the potential function

$$\frac{2k^{1/d}\sum_{F\in\mathcal{F}}\pi(F)}{|X|^{1/d}}, \tag{6}$$

which is slightly more restrictive than the one derived from Theorem 2.1. In the proof of Theorem 3.1, we apply successively Theorem 2.1 with decreasing $t$ at each iteration which gives potential function: $i \cdot \frac{2k^{\frac{1}{d}}\sum_{F\in\mathcal{F}}\pi(F)}{|X|^{\frac{1}{d}}}$ for the elements of the $i^{th}$ part. In our experiments, we use a potential function that does not depend on the number of parts already built. That is, the two potential functions are equal while building the first part but the experimental potential function remains constant when the theoretical one increases linearly after each part is built.

We also exclude some implementation details in the pseudo-code of GreedyPotential. For instance, after selecting $y_k$, we do not immediately update $\omega(x)$ with all ranges in $\mathcal{C}(P_i, y_k)$. We store these ranges in a queue and only update $\omega(x)$ in the next iteration, range by range, until we can find an element within the potential function rate for the next iteration.

The algorithmic bottleneck of GreedyPotential is the weight update operation (i.e. updating $\omega(\cdot)$) which gives an overall time complexity of $O(nmt)$, since it is only performed at most once per partition for each pair $(x, F) \in X \times \mathcal{F}$.

## 4 Variants

**Min Weight** In this variant of GreedyPotential, instead of selecting an arbitrary element satisfying the upper bound of the potential function, we pick the element with the lowest weight at the time. This variant has the same asymptotic complexity of $O(nmt)$ as GreedyPotential. The pseudocode of MinWeight is in the Appendix.

As we will see later, our experiments show that MinWeight generally finds partitions with lower crossing numbers and runs faster than GreedyPotential. It is beneficial to spend some extra time to search for the vertex with *the lowest* $\omega(\cdot)$ value at every iteration, as it then decreases the total number of crossings.

**Part At Once** Next, to improve the running time, we present a different approach where weight updates are done only when a part of the partition has been built. This has the added benefit that this can easily be parallelized.

---

**Algorithm 2:** PartAtOnce

---

1  $n \leftarrow |X|, m \leftarrow |\mathcal{F}|, \mathcal{P} \leftarrow \emptyset$
2  $\forall F \in \mathcal{F}, \pi(F) \leftarrow 1$
3  **for** $i \leftarrow 1$ to $t$ **do**
4      $x_0 \leftarrow$ a random element of $X$
5      $\forall x \in X, \omega(x) \leftarrow 0$
6      **for** $w$ steps **do**
7          $S \leftarrow$ random range with $\forall f \in \mathcal{F}, P(S = F) = \frac{\pi(F)}{\sum_{G \in \mathcal{F}} \pi(G)}$
8          **foreach** $x \in X$ such that $S$ crosses $\{x, x_0\}$ **do**
9              $\omega(x) \leftarrow \omega(x) + \pi(S)$
10     $P_i \leftarrow \{x_0\} \cup M_\omega$ where $M_\omega$ contains the $\frac{n}{t} - 1$ elements of $X$ with the smallest weight w.r.t $\omega$
11     $\forall F \in \mathcal{F}, \pi(F) \leftarrow \pi(F) \cdot 2^{I(P_i, F)}$
12     $\mathcal{P} \leftarrow \mathcal{P} \cup \{P_i\}$
13     $X \leftarrow X \setminus P_i$
14 **return** $\mathcal{P}$

---

This algorithm builds each part of the partition by picking a random first element $x_0$ and then adding the $\frac{n}{t} - 1$ elements with the lowest $\omega$ values. Once a part is built, we update the weights of the ranges using the MWU rule as before. Furthermore, we only use an approximation of the cost of adding an element: we sample $w = \Theta(t)$ ranges according to their weights and compute elements' cost only with respect to these ranges. The resulting algorithm is given in PartAtOnce. The overall time complexity is $O(mn + twn)$, as the weight update takes time $O(\frac{mn}{t})$. Another advantage of this algorithm is that the weight update step now can be parallelized to as many as $m$ cores.

# 5 Experiments

We now turn to an evaluation of our algorithms on a variety of data.

Our initial experiments, including the ones for parameter setting are performed on geometric set systems induced by half-spaces in $\mathbb{R}^d$. There are two reasons for this:

**Lower bound.** The base set $X$ can be constructed such that any partition into $t$ parts has crossing number $\Omega\left(t^{1-\frac{1}{d}}\right)$.

**Comparability.** The only known implementation of low-crossing partitions is for half-space set systems in $\mathbb{R}^2$ [MP18].

In particular, we consider the following set system $(X, \mathcal{F})$.

**Grid set system.** Given two parameters $n, d \in \mathbb{N}$, the base set $X$ consists of $n$ points in the unit hypercube $[0, 1]^d$ uniformly at random (each coordinate is set independently and uniformly). The range set $\mathcal{F}$ is induced by $dn^{1/d}$ grid-like halfspaces (for each of the standard basis vectors, we add $n^{1/d}$ evenly-spaced half-spaces orthogonal to this vector, each containing the point $(1, ..., 1)$).[5]

Regarding non-geometric set systems, we run our algorithms on different types of graphs (power-law neighborhood hypergraphs, Facebook social circles data, and ArXiv co-authorship data from the SNAP dataset collection [LK14]). We also provide in Section 5.3.3 experiments on lines in the projective planes where partitions have been proved to have linear crossing number for $t = O(\sqrt{n})$ [AHW87] and we obtain results in agreement to this.

## 5.1 Implementations

We provide two implementations for the partition algorithms, one in C++ and one in Rust. The Rust implementation is faster but more memory consuming. They are compatible as they use the same file format to store set systems and partitions. The Rust implementation does not implement all types of set systems generation as the set systems can be generated efficiently with the C++ implementation and partitioned afterwards with the Rust code. We also did not reimplement some evaluation and experiment functions such as violation number computation in Rust. The Rust implementation should mainly be used to quickly partition set systems stored in files. The runtimes we give in this work have been obtained with the Rust implementation.

We ran our experiments on a home computer with 16GB of RAM and an AMD Ryzen 7 5800X (16 cores) @ 4.85 GHz, all 16 cores were used for PartAtOnce.

---

[5]We also tested our algorithms on the variants of this set system with non-uniform point distributions and other types of halfspaces distribution such as random halfspaces (data available in the Appendix) and non-uniform grids. We obtained similar results in terms of crossing number on non-uniform halfspaces-spanned set systems and therefore did not include these results in this work.

## 5.2 Algorithm Parameters

### 5.2.1 Experiments on the number of samples to approximate the weights in PartAtOnce

Recall that the algorithm PartAtOnce has an input parameter $w$ which determines how many weight-updates we do each iteration. That is, we want to estimate the sums $\sum_{F \in \mathcal{F}} \pi(F) I(F, \{x_0, x\})$ for all $x \in X$.

We study the influence of the parameters of set systems on the number of samples required to obtain a good approximation of the weight of each element. We draw in Figure 3 the evolution of the crossing number depending on $w$ with varying parameters $n, d, t$. This experiment has been obtained by averaging 10 runs of PartAtOnce on grid set systems with $n = 8192, d = 2$ and $t = 128$ when they are not the varying parameter.

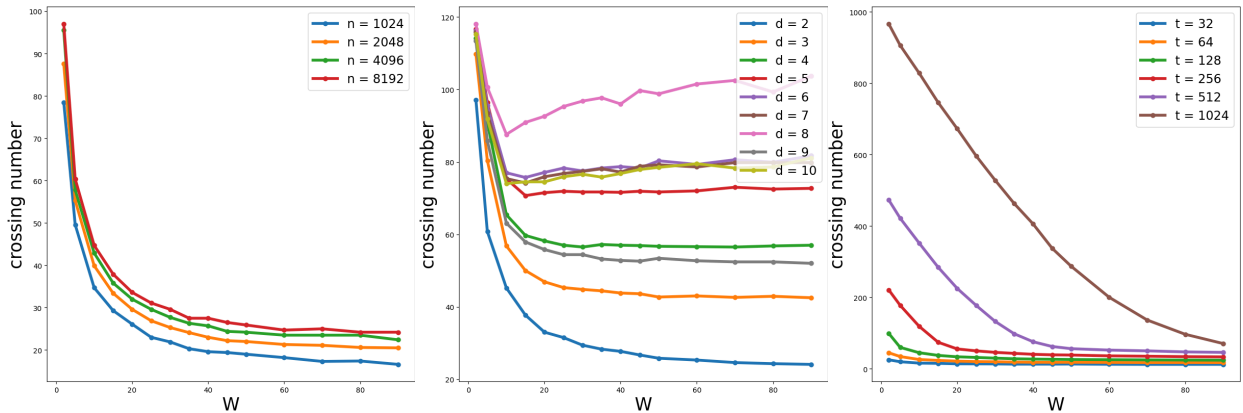

Figure 3: Evolution of the crossing number with $w$ on grid set systems with varying $n, d, t$ from left to right

As we can see on the figures, increasing $w$ will decrease the crossing number. However after some point the decrease in the crossing number is small. At this point, further increasing $w$ is not interesting as it increase runtime without significantly improving the crossing number.

We see that changing $n$ and $d$ does not change the number of samples required to obtain a small crossing number. The third graph reveals that the number of sample should be a function of $t$ and interpolating the results gives that $w \approx \frac{t}{2}$ is the point where the crossing number decrease is slow.

Thus we set $w = \max(30, \frac{t}{2})$ for all the experiments with PartAtOnce on the grid set system.

We repeated this experiment on random halfspaces and random power-law graph set systems. We obtained that the number of samples to estimate well the element's weight similarly only depends on $t$. However, the number of samples to obtain a small crossing w.r.t. $n$ and $d$ is higher, therefore, we set $w = \max(100, \frac{t}{2})$ for the other experiments.

### 5.2.2 Number of potential function violations

We have shown that the overall crossing number is bounded if we always extend parts with elements such that $\text{cost}(P_i) + \omega(x)$ is below the potential function of Equation 3.

In GreedyPotential and MinWeight, we keep track of the number of iterations where there was no element such that $\text{cost}(P_i) + \omega(x)$ is below the upper bound provided by the potential function with $\text{cost}(P_i)$ the weight of the partial part (i.e. sum of the elements' weight added to it) during the course of the algorithm.

As we noted before, the potential function used in our implementations is stricter than the one provided by Theorem 2.1 (the theoretical potential is larger on all but the first part). We study whether it violates the experimental potential function as well as the theoretical potential function from Theorem 2.1.

Figure 4 visualizes how the number of violations evolve during the construction of each part of the partition. It consists of blocks of 4 lines, each block containing the data for 8192 points on the $2-$dimensional grid set system with, from top to bottom, 16, 32 and 64 parts. The 4 lines correspond (from top to bottom) to the following algorithms and violation measures:

1. GreedyPotential , with violations according to the potential function of Theorem 2.1,
2. MinWeight, with violations according to the potential function of Theorem 2.1,
3. GreedyPotential, with violations according to the potential function of Equation 6,
4. MinWeight, with violations according to the potential function Equation 6.

Each line consists of $t$ many squares, where the color of the $i^{th}$ square encodes the number of violations occurred during the construction of the $i^{th}$ part.

These data have been obtained by averaging over the results of 100 runs of the algorithms. We observe that MinWeight rarely picks an element violating the potential function bound, even with respect to the practical, stricter potential function. However, when looking at GreedyPotential compared to the experimental potential function, we see an increase in the number of violations. Interestingly, in most of the cases, we still stay below the theoretical potential function bound.

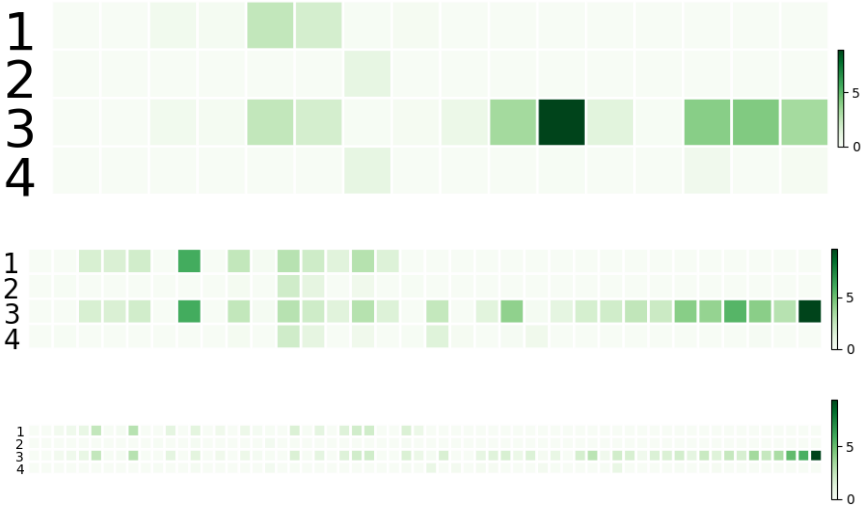

Figure 4: Number of potential function bound violations.

For reference, we also include numerical data on the number of potential function violations in Table 1 with bigger partition sizes.

| input n,d,t | GreedyPotential # violations | MinWeight # violations |
|---|---|---|
| 8192,2,128 | 159.4 | 2.1 |
| 8192,2,256 | 173.4 | 4.3 |
| 8192,2,512 | 121.9 | 1.1 |
| 8192,2,1024 | 84.0 | 21.3 |
| 8192,2,2048 | 60.9 | 11.6 |
| 8192,2,512 | 121.9 | 1.1 |
| 8192,3,512 | 146.7 | 47.2 |
| 8192,4,512 | 141.6 | 47.6 |
| 8192,5,512 | 96.4 | 33.5 |
| 8192,10,512 | 1.7 | 16.5 |
| 2048,2,512 | 20.9 | 0 |
| 4096,2,512 | 49.5 | 0 |
| 8192,2,512 | 121.9 | 0 |
| 16384,2,512 | 232.7 | 7.0 |

## 5.3 Performance evaluation

### 5.3.1 Grid set system

The grid set system has two parameters, $n, d \in \mathbb{N}$ and is constructed as follows. We take $n$ points in the unit hypercube $[0, 1]^d$, picking each coordinate independently and uniformly. For each of the standard basis vectors, add $n^{\frac{1}{d}}$ evenly-spaced halfspaces orthogonal to this vector, each containing the point $(1, ..., 1)$.

We studied the evolution of both the crossing number and runtime of our algorithms depending on the different variables $n, d$ and $t$ of the grid set system. Figure 5 illustrates the results of the algorithms averaged over 10 executions. We include the raw data of experiments on grid set systems and random halfspaces in Table 6 in the Appendix

For $d = 2$, we compare our method to **MP-Matoušek**, which is the implementation of Matoušek's algorithm by Matheny and Phillips [MP18] in Python[6]. **MP-Matoušek** uses the branch factor of the polytree they build to construct cuttings as an optimization parameter: increasing it can reduce the crossing number but also increases runtime. We use the default branching factor provided by their implementation in our experiments.

---

[6] The code of Matheny is available on Github [Mat18], we modified it to be able to use it on our input data and made the modifications available on Github.

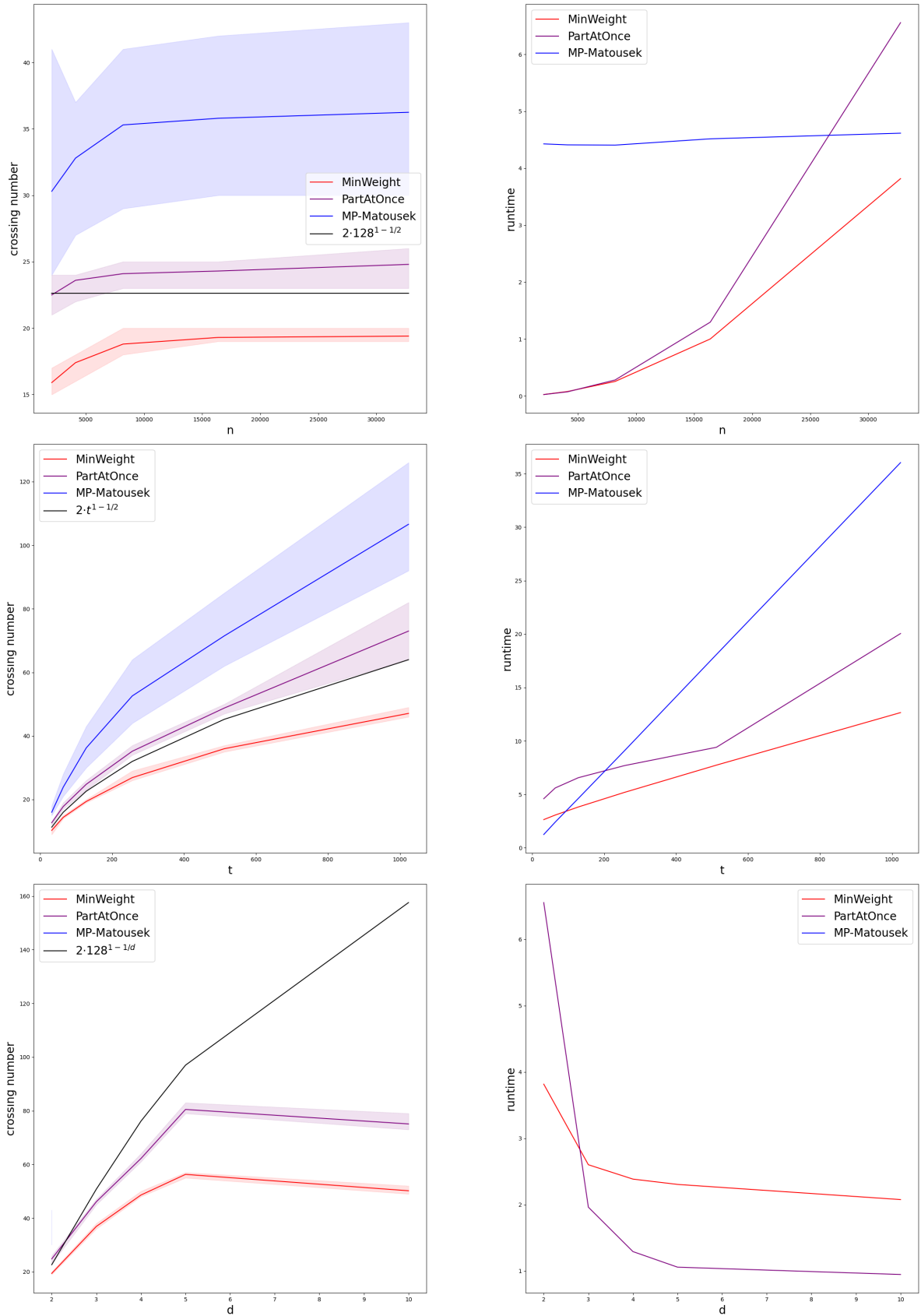

Figure 5: Average crossing numbers and runtimes of the 3 variations of the algorithm depending on the parameters $n, d, t$ on the grid set system. The curves trace the averages, the shaded area corresponds to $\pm 1$ standard deviation. Each parameter has been tested independently and we set $n = 8192$, $d = 2$ and $t = 512$ when they are not the parameter varying.

The results are presented in Figure 5. The left column represents the crossing numbers with, from top to bottom, varying $n$, $d$, and finally $t$. The lines represent $\kappa_{\mathcal{F}}$. The black line marks $2t^{1-\frac{1}{d}}$, that is the order of the crossing number that can be achieved for set systems induced by halfspaces. The red lines corresponds to MinWeight, the purple one to PartAtOnce and the blue lines to **MP-Matoušek**. The graphs on the right represent the runtimes with the same color code.

We see that MinWeight and PartAtOnce obtain crossing numbers close to the optimal bounds of $t^{1-\frac{1}{d}}$. PartAtOnce is significantly faster than MinWeight on large set systems even if on small ones the overhead of parallelization makes it slower. This is particularly visible on random halfspaces set system partition as $|\mathcal{F}|$ is larger (cf Table 6). Since GreedyPotential was performing worse than the other two methods on abstract systems as well, we omitted further data on this algorithm for readability purpose but it is included for reference in Table 6 in the Appendix.

We even see that MinWeight and PartAtOnce consistently outperform **MP-Matoušek** in terms of crossing number. However, for large halfspaces-spanned set systems, **MP-Matoušek** is faster than our algorithms as it uses ranges sampling to compute cuttings (cf. random halfspaces results in Table 6). Sampling also allows their implementation to maintain a stable runtime when $n$ and $m$ increases when our runtimes increases with both. However our implementations' runtime doesn't increase as much when $t$ increases. We also tried adding ranges sampling in our weight computation, the results we obtained were not conclusive.

### 5.3.2 Abstract set systems induced by neighborhoods in graphs

Now we turn to experiments on abstract set systems. We will focus on set systems induced by closed neighborhoods in graphs, which allows us to perform experiments both on large-scale synthetic data (power-law random graphs), and on real-word network datasets. Given a graph $G = (V, E)$, the neighborhood set system of $G$ is a set system $(X, \mathcal{F})$ with $X = V$, and $\mathcal{F}$ consists of closed neighborhoods of vertices, that is, $\mathcal{F} = \{\{y \in V : (x, y) \in E\} \cup \{x\} : x \in V\}$. Our implementation initiates the study of low-crossing partitions for these set systems; as there are no theoretical guarantee for their existence, we compare our results with $t^{1-\frac{1}{d}}$ where $d$ is the VC-dimension.[7]

**Power-law random graphs.** A random graph generated with respect to the power-law distribution $(n, \beta)$ is a graph where the probability for a vertex to have degree $c \geq 1$ is proportional to $\frac{1}{c^{\beta}}$ [ACL01]. Coudert et al. [Cou+24] studied the expected VC-dimension of power-law graphs. We use their average observed VC-dimension to evaluate the crossing numbers obtained by our algorithms. The results are presented in Table 2.

---

[7]Note that the neighborhood set system of a graph is self-dual, thus $d$ is equal to the dual VC-dimension as well.

| input n,β,t | VC-dim [Cou+24] | $t^{1-1/d}$ | MinWeight | | PartAtOnce | |
|---|---|---|---|---|---|---|
| | | | $\kappa_{\mathcal{F}}$ | runtime (s) | $\kappa_{\mathcal{F}}$ | runtime (s) |
| 2000,2,32 | 5.2 | 16.43 | 10.4 | 0.0139 | 18.05 | 0.00837 |
| 2000,2.5,32 | 3.8 | 12.85 | 8.0 | 0.0132 | 15.5 | 0.00789 |
| 2000,3,32 | 3 | 10.08 | 6.1 | 0.0128 | 13.5 | 0.00762 |
| 2000,2,128 | 5.2 | 50.35 | 17.25 | 0.0206 | 28.5 | 0.0251 |
| 2000,2.5,128 | 3.8 | 35.7 | 10.3 | 0.0181 | 20.4 | 0.0237 |
| 2000,3,128 | 3 | 25.4 | 6.8 | 0.017 | 15.8 | 0.0227 |
| 2000,2,512 | 5.2 | 154.3 | 18.1 | 0.0398 | 31.35 | 0.0908 |
| 2000,2.5,512 | 3.8 | 99.15 | 10.1 | 0.0335 | 20.2 | 0.0865 |
| 2000,3,512 | 3 | 64.0 | 6.7 | 0.0293 | 14.8 | 0.0824 |
| 4000,2,32 | 5.8 | 17.61 | 10.95 | 0.0682 | 19.9 | 0.0171 |
| 4000,2.5,32 | 4.05 | 13.6 | 8.0 | 0.0682 | 17.1 | 0.0157 |
| 4000,3,32 | 3 | 10.08 | 6.3 | 0.0664 | 14.4 | 0.0152 |
| 4000,2,128 | 5.8 | 55.45 | 20.95 | 0.0845 | 34.55 | 0.0524 |
| 4000,2.5,128 | 4.05 | 38.63 | 12.2 | 0.0796 | 24.6 | 0.0478 |
| 4000,3,128 | 3 | 25.4 | 8.2 | 0.0812 | 18.1 | 0.0461 |
| 4000,2,512 | 5.8 | 174.6 | 26.4 | 0.145 | 51.0 | 0.189 |
| 4000,2.5,512 | 4.05 | 109.7 | 12.3 | 0.127 | 26.5 | 0.174 |
| 4000,3,512 | 3 | 64.0 | 8.1 | 0.116 | 18.2 | 0.169 |
| 30000,2,32 | 6.8 | 19.22 | 11.0 | 11.1 | 23.8 | 0.782 |
| 30000,2.5,32 | 4.75 | 15.43 | 8.2 | 11.1 | 21.2 | 0.727 |
| 30000,3,32 | 3 | 10.08 | 6.2 | 11.2 | 17.6 | 0.707 |
| 30000,2,128 | 6.8 | 62.71 | 24.3 | 12.0 | 51.7 | 1.52 |
| 30000,2.5,128 | 4.75 | 46.09 | 13.2 | 11.8 | 37.4 | 1.42 |
| 30000,3,128 | 3 | 25.4 | 8.8 | 11.8 | 24.2 | 1.35 |
| 30000,2,512 | 6.8 | 204.6 | 49.2 | 13.5 | 175.8 | 3.6 |
| 30000,2.5,512 | 4.75 | 137.7 | 18.0 | 13.2 | 51.2 | 3.54 |
| 30000,3,512 | 3 | 64.0 | 9.2 | 12.9 | 28.8 | 3.44 |

Table 2: Crossing number and runtime of our algorithms on the power-law graph neighborhood set system.

The crossing numbers that our algorithms obtain are comparable to $t^{1-\frac{1}{d}}$. This suggests that low-crossing partitions might exist in abstract set systems. Similarly to the geometric case, MinWeight obtains the best crossing number among the different variants and the best runtime is obtained with PartAtOnce due to its parallel execution.

Finally, we tested our algorithms on two *real world* network datasets.

**Facebook social circles.** We ran our algorithms on set systems generated from a graph representing the friendships between Facebook users. We use the data set from McAuley and Leskovec [ML12], which is composed of the network induced by the friends of 10 users. The graph is composed of 4039 nodes and 88234 edges. The VC-dimension of this graph is 6 [Cou+24]. Given a graph $G = (V, E)$ and a radius $r \in \mathbb{N}$, we now study the set system with base set $V$, where each vertex $v \in V$ defines a range containing all elements at distance at most $r$ from $v$ in $G$, that is, our set of ranges is defined as $\mathcal{F} = \{\{y \in V : \text{dist}_G(x, y) \leq r\} : x \in V\}$. We compile the results obtained for experiments on social network data in Table 3.

| input | MinWeight | | PartAtOnce | |
| r,t | $\kappa_\mathcal{F}$ | runtime (s) | $\kappa_\mathcal{F}$ | runtime (s) |
| --- | --- | --- | --- | --- |
| 1,10 | 6 | 0.0677 | 10 | 0.0213 |
| 1,20 | 7 | 0.0709 | 14 | 0.0286 |
| 1,40 | 9 | 0.0949 | 17 | 0.0364 |
| 2,10 | 7 | 0.115 | 7 | 0.324 |
| 2,20 | 6 | 0.188 | 10 | 0.431 |
| 2,40 | 10 | 0.334 | 14 | 0.468 |
| 3,10 | 6 | 0.146 | 6 | 0.487 |
| 3,20 | 7 | 0.247 | 8 | 0.618 |
| 3,40 | 9 | 0.422 | 9 | 0.683 |

Table 3: Crossing number and runtime of our algorithms on the power-law graph neighborhood set system.

The graph upon which we build our set system is relatively sparse: the average degree is 21.84, but the maximum degree is 1045. We observe that our crossing numbers are consistently below the $t^{1-\frac{1}{d}}$ bound, where $d$ is the VC-dimension of the graph.

**ArXiv co-authorship graph.** We also did experiments on the neighborhood set system of the collaboration graph from the High Energy Physics - Phenomenology arXiV subject [LKF07]. It is composed of 12008 nodes and 118521 edges, with a maximum degree of 491 and VC-dimension 5 [Cou+24]. We ran experiments with higher values of $t$ and observe that the crossing numbers remain low.

| input r,t | MinWeight $\kappa_{\mathcal{F}}$ | MinWeight runtime (s) | PartAtOnce $\kappa_{\mathcal{F}}$ | PartAtOnce runtime (s) |
|---|---|---|---|---|
| 1,50 | 12 | 0.973 | 21 | 0.182 |
| 1,100 | 16 | 1.08 | 28 | 0.256 |
| 1,200 | 19 | 1.17 | 34 | 0.373 |
| 1,500 | 26 | 1.46 | 41 | 0.74 |
| 2,50 | 30 | 1.6 | 42 | 0.885 |
| 2,100 | 48 | 2.3 | 69 | 1.03 |
| 2,200 | 73 | 3.8 | 118 | 1.42 |
| 2,500 | 121 | 7.29 | 254 | 1.94 |
| 3,50 | 41 | 4.04 | 49 | 2.42 |
| 3,100 | 72 | 7.01 | 90 | 3.0 |
| 3,200 | 126 | 12.8 | 172 | 4.06 |
| 3,500 | 287 | 29.3 | 384 | 5.71 |

Table 4: Crossing number and runtime for our algorithms on the ArXiv co-authorship graph.

### 5.3.3 Finite projective planes

Projective planes of order $a \in \mathbb{N}$ are a set of points $X$ and lines $\mathcal{F}$ with the following properties:

- $n = |X| = |\mathcal{F}| = a^2 + a + 1$
- $\forall x \in X, |\{F \in \mathcal{F} : x \in F\}| = a + 1$
- $\forall F \in \mathcal{F}, |F \cap X| = a + 1$

Alon, Haussler and Welzl [AHW87] showed that for $t = O(\sqrt{n})$, finite projective planes do not admit partitions of size $t$ with sublinear crossing number even though these set systems have VC-dimension 2. We tested our algorithms on the embedding of projective planes in dimension 3, that is $X \subseteq \mathbb{N}^3$ and lines are constructed to meet properties 2 and 3. We only implemented this sets system for prime orders as this embedding does not work with some non-primes order. The data for our experiments on projective planes of order 233 is in Table 5.

| input n,t | PartAtOnce $\kappa_F$ |
|---|---|
| 54523,200 | 185 |
| 54523,500 | 229 |
| 54523,1000 | 234 |
| 54523,2000 | 234 |

Table 5: Crossing number and runtime of our algorithms on the power-law graph neighborhood set system.

As expected our algorithms obtains the maximum intersection number for high number of partitions.

### 5.3.4 $\varepsilon$-Approximations

Simplicial partitions are a way to compute $\varepsilon$-approximation, defined as:

**Definition 5.1. $\varepsilon$-approximation** Let $(X, \mathcal{F})$ be a set system. Then a set $A \subseteq X$ is an $\varepsilon$-approximation iff

$$\forall F \in \mathcal{F}, \left| \frac{|F|}{|X|} - \frac{|F \cap A|}{|A|} \right| \leq \varepsilon$$

[LLS01, Tal94] showed that a uniform random sample of size $\Theta(\frac{d}{\varepsilon^2})$ where $d$ is the VC-dimension of $(X, \mathcal{F})$ is an $\varepsilon$-approximation with high probability. For comparison, we compute a low-crossing partition and pick one element uniformly at random from each part created (see [Mus22] for more details).

**Lemma 5.2.** [STZ06] Let $\{P_1, ..., P_t\}$ be a $\left(t, t^{1-\frac{1}{d}}\right)$-partition of a sets system $(X, \mathcal{F})$, then a set $A = \{x_1, ..., x_r\} \subseteq X$ such that $x_1, ..., x_r$ are selected uniformly at random respectively in $P_1, ..., P_r$ is a $\tilde{O}\left(\left(\frac{1}{t}\right)^{\frac{d+1}{2d}}\right)$-approximation of $(X, \mathcal{F})$ with constant probability.

This result relies on the fact that for a given set $F \in \mathcal{F}$, a part that does not intersect $F$ does not contribute to the discrepancy between $|F|$ and $|F \cap A|$. The proof then proceeds by upper-bounding the discrepancy using a concentration inequality over the—at most $t^{1-\frac{1}{d}}$—parts intersecting $F$.

We compare the error factor $\varepsilon = \max_{F \in \mathcal{F}} \left| \frac{|F|}{|X|} - \frac{|F \cap A|}{|A|} \right|$ obtained with random sample of size $t$ as well as those obtained with an approximation constructed from a $t$-partition on our different algorithms over the grid set system. The results are compiled in Figure 6. Talagrand [Tal94] showed that a uniform random sample of size $t$ is an $O\left(\frac{1}{\sqrt{t}}\right)$-approximation.

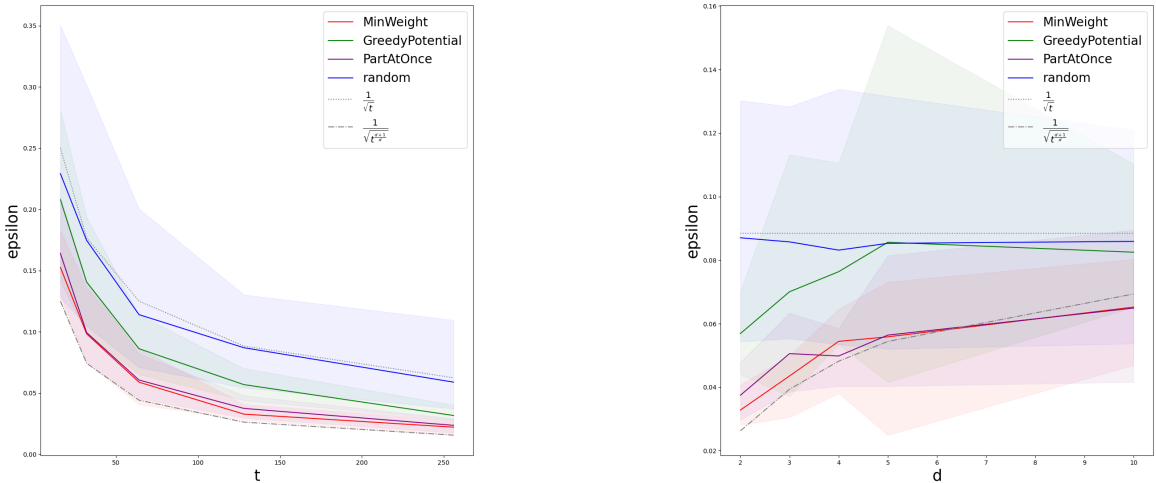

Figure 6: Value of the error factor $\varepsilon$ for varying $t$ on the 2-dimensional grid set system on top and $d$ on 64 parts of the $d$-dimensional grid set system on the bottom. The grey curves represent $\frac{1}{\sqrt{t}}$ and $\frac{1}{\sqrt{t^{\frac{d+1}{d}}}}$ which are respectively the expected error factor for a uniform random sample of size $t$ and the optimal error factor for the $d$-dimensional grid sets system

As we can see, our algorithms outperform the random sample (in blue) and, in particular, MinWeight and GreedyPotential obtain an error factor close to the optimal one $\frac{1}{\sqrt{t^{\frac{d+1}{d}}}}$ [Ale90] (see the known bounds section of [Mus22] for a sketch of the proof). The raw data for the experiment is available in Table 7 in the Appendix.

# Bibliography

[ACL01] Aiello, William ; Chung, Fan ; Lu, Linyuan: A Random Graph Model for Power Law Graphs. In: *Experimental Mathematics* vol. 10 (2001)

[AHW87] Alon, N. ; Haussler, D. ; Welzl, E.: Partitioning and geometric embedding of range spaces of finite Vapnik-Chervonenkis dimension.. In: *Proceedings of the third annual symposium on Computational geometry*, 1987

[Ale90] Alexander, Ralph: Geometric methods in the study of irregularities of distribution. In: *Combinatorica* vol. 10 (1990)

[AMS13] Agarwal, Pankaj K. ; Matoušek, Jiří ; Sharir, Micha: On Range Searching with Semialgebraic Sets II. In: *SIAM Journal on Computing* vol. 42 (2013)

[Cha12] Chan, Timothy M.: Optimal Partition Trees. In: *Discrete & Computational Geometry* vol. 47 (2012)

[Cha93] Chazelle, B.: Cutting hyperplanes for divide-and-conquer. In: *Discrete & Computational Geometry* vol. 9 (1993)

[Cou+24] Coudert, David ; Csikós, Mónika ; Ducoffe, Guillaume ; Viennot, Laurent: Practical Computation of Graph VC-Dimension.. In: *Symposium on Experimental Algorithms (SEA)*, 2024

[Har00] Har-Peled, S.: Constructing Planar Cuttings in Theory and Practice. In: *SIAM J. Comput.* vol. 29 (2000)

[LK14] Leskovec, Jure ; Krevl, Andrej: SNAP Datasets: Stanford Large Network Dataset Collection. In: *http://snap.stanford.edu/data* (2014)

[LKF07] Leskovec, Jure ; Kleinberg, Jon ; Faloutsos, Christos: Graph evolution: Densification and shrinking diameters. In: *ACM transactions on Knowledge Discovery from Data (TKDD)* vol. 1 (2007)

[LLS01] Li, Yi ; Long, Philip M. ; Srinivasan, Aravind: Improved Bounds on the Sample Complexity of Learning. In: *Journal of Computer and System Sciences* vol. 62 (2001)

[Mat18] Matheny, Michael: pypartition. In: *https://github.com/michaelmathen/pypartition*, GitHub (2018)

[Mat92] Matoušek, Jiří: Efficient partition trees. In: *Discrete & Computational Geometry* vol. 8 (1992)

[Mat93] Matoušek, Jiří: Range searching with efficient hierarchical cuttings. In: *Discrete & Computational Geometry* vol. 10 (1993)

[ML12] McAuley, Julian ; Leskovec, Jure: Learning to discover social circles in ego networks.. In: *Proceedings of the 25th International Conference on Neural Information Processing Systems - Volume 1, NIPS'12*, 2012

[MP18] Matheny, Michael ; Phillips, Jeff M: Practical Low-Dimensional Halfspace Range Space Sampling.. In: *26th Annual European Symposium on Algorithms (ESA)*, 2018

[Mus22] Mustafa, Nabil H: *Sampling in combinatorial and geometric set systems.* vol. 265 : American Mathematical Society, 2022

[STZ06]   Suri, Subhash ; Toth, Csaba D ; Zhou, Yunhong: Range Counting over Multidimensional Data Streams. In: *Discrete & Computational Geometry* vol. 36 (2006)

[Tal94]   Talagrand, M.: Sharper Bounds for Gaussian and Empirical Processes. In: *The Annals of Probability* vol. 22 (1994)

# 6 Appendix

**Algorithm 3:** MinWeight

1. $n \leftarrow |X|, m \leftarrow |\mathcal{F}|, \mathcal{P} \leftarrow \emptyset$
2. $\forall F \in \mathcal{F}, \pi(F) \leftarrow 1$
3. **for** $i \leftarrow 1$ to $t$ **do**
4.     $x_0 \leftarrow$ a random element of $X$
5.     $P_i \leftarrow \{x_0\}$
6.     cost $\leftarrow 0$
7.     **for** $F \in \mathcal{F}$ **do**
8.         **foreach** $x \in X$ with $F$ crossing $\{x_0, x\}$ **do**
9.             $\omega(x) \leftarrow \omega(x) + \pi(F)$
10.     **for** $k \leftarrow 2$ to $\frac{n}{t}$ **do**
11.         $y_k \leftarrow \operatorname{argmin}_{x \in X} \omega(x)$
12.         $X \leftarrow X \setminus \{y_k\}$
13.         cost $\leftarrow$ cost $+ \omega(y_k)$
14.         **foreach** $F \in \mathcal{C}(P_i, y_k)$ **do**
15.             **foreach** $x \in X$ with $F$ crossing $\{x_0, x\}$ **do**
16.                 $\omega(x) \leftarrow \omega(x) - \pi(F)$
17.         $P_i \leftarrow P_i \cup \{y_k\}$
18.     $\forall F \in \mathcal{F}, \pi(F) \leftarrow \pi(F) \cdot 2^{I(P_i, F)}$
19.     $\mathcal{P} \leftarrow \mathcal{P} \cup \{P_i\}$
20.     $X \leftarrow X \setminus P_i$
21. **return** $\mathcal{P}$

| input n,d,t | $2t^{1-1/d}$ | **MP-Matoušek** $\kappa_\mathcal{F}$ | runtime (s) | MinWeight $\kappa_\mathcal{F}$ | runtime (s) | GreedyPotential $\kappa_\mathcal{F}$ | runtime (s) | PartAtOnce $\kappa_\mathcal{F}$ | runtime (s) |
|---|---|---|---|---|---|---|---|---|---|
| | | | | Grid | | | | | |
| 2048,2,128 | 22.6 | 30.3 | 4.43 | 15.9 | 0.0265 | 45.4 | 0.029 | 22.5 | 0.0253 |
| 4096,2,128 | 22.6 | 32.8 | 4.41 | 17.4 | 0.0789 | 50.0 | 0.0867 | 23.6 | 0.0734 |
| 8192,2,128 | 22.6 | 35.3 | 4.41 | 18.8 | 0.256 | 49.0 | 0.259 | 24.1 | 0.281 |
| 16384,2,128 | 22.6 | 35.8 | 4.52 | 19.3 | 1.0 | 52.3 | 0.848 | 24.3 | 1.3 |
| 32768,2,128 | 22.6 | 36.25 | 4.62 | 19.4 | 3.82 | 50.8 | 2.83 | 24.8 | 6.56 |
| 32768,2,32 | 11.3 | 16.0 | 1.24 | 10.3 | 2.64 | 22.6 | 1.59 | 12.7 | 4.59 |
| 32768,2,64 | 16.0 | 23.9 | 2.41 | 14.4 | 3.06 | 34.5 | 1.97 | 17.8 | 5.6 |
| 32768,2,128 | 22.6 | 36.25 | 4.62 | 19.4 | 3.82 | 50.8 | 2.83 | 24.8 | 6.56 |
| 32768,2,256 | 32.0 | 52.6 | 9.07 | 26.9 | 5.18 | 99.8 | 4.3 | 35.2 | 7.69 |
| 32768,2,512 | 45.3 | 71.5 | 18.1 | 36.0 | 7.74 | 211.8 | 7.45 | 48.8 | 9.4 |
| 32768,2,1024 | 64.0 | 106.6 | 36.0 | 47.1 | 12.7 | 566.6 | 14.7 | 73.0 | 20.1 |
| 32768,2,128 | 22.6 | 36.25 | 4.62 | 19.4 | 3.82 | 50.8 | 2.83 | 24.8 | 6.56 |
| 32768,3,128 | 50.8 | | | 36.9 | 2.6 | 91.1 | 1.12 | 46.1 | 1.96 |
| 32768,4,128 | 76.1 | | | 48.6 | 2.38 | 104.2 | 0.855 | 62.2 | 1.29 |
| 32768,5,128 | 97.0 | | | 56.3 | 2.3 | 128.0 | 0.657 | 80.5 | 1.06 |
| 32768,10,128 | 158 | | | 50.2 | 2.08 | 128.0 | 0.311 | 75.1 | 0.946 |
| | | | | Random Halfspaces | | | | | |
| 2048,2,128 | 22.6 | 34.5 | 4.3 | 19.0 | 2.8 | 49.5 | 3.27 | 27.8 | 0.764 |
| 4096,2,128 | 22.6 | 37.2 | 4.35 | 20.5 | 12.0 | 54.4 | 13.9 | 28.7 | 5.52 |
| 8192,2,128 | 22.6 | 40.0 | 4.43 | 21.9 | 50.1 | 53.2 | 56.7 | 28.4 | 45.1 |
| 8192,2,32 | 11.3 | 17.6 | 1.13 | 12.5 | 18.0 | 24.6 | 19.6 | 14.2 | 34.6 |
| 8192,2,64 | 16.0 | 26.3 | 2.26 | 16.3 | 29.5 | 38.1 | 33.4 | 20.3 | 40.3 |
| 8192,2,128 | 22.6 | 40.0 | 4.43 | 21.9 | 50.1 | 53.2 | 56.7 | 28.4 | 45.1 |
| 8192,2,256 | 32.0 | 55.3 | 9.0 | 28.4 | 89.6 | 110.0 | 108 | 40.4 | 49.1 |
| 8192,2,512 | 45.3 | 79.5 | 17.6 | 36.2 | 166 | 240.4 | 214 | 46.3 | 61.8 |
| 8192,2,128 | 22.6 | 40.0 | 4.43 | 21.9 | 50.1 | 53.2 | 56.7 | 28.4 | 45.1 |
| 8192,3,128 | 50.8 | | | 44.5 | 66.1 | 92.8 | 76.9 | 55.9 | 37.4 |
| 8192,4,128 | 76.1 | | | 64.7 | 81.5 | 113.7 | 98.5 | 86.6 | 27.9 |
| 8192,5,128 | 97.0 | | | 82.3 | 94.4 | 122.4 | 86.7 | 107.0 | 20.3 |
| 8192,10,128 | 158 | | | 121.3 | 128 | 128.0 | 90.5 | 128.0 | 10.4 |

Table 6: $\kappa_F$ and runtime of our algorithms on the grid set system and a set system generated by halfspaces.

| input n,d,t | Random Sample Error Factor | MinWeight Error Factor | GreedyPotential Error Factor | PartAtOnce Error Factor |
|---|---|---|---|---|
| 8192,2,16 | 0.2292 | 0.1526 | 0.2078 | 0.164 |
| 8192,2,32 | 0.1747 | 0.09839 | 0.1407 | 0.0994 |
| 8192,2,64 | 0.114 | 0.05889 | 0.08621 | 0.06064 |
| 8192,2,128 | 0.08704 | 0.03281 | 0.0569 | 0.03748 |
| 8192,2,256 | 0.05892 | 0.02216 | 0.0317 | 0.02366 |
| 8192,3,16 | 0.2463 | 0.1723 | 0.2419 | 0.1819 |
| 8192,3,32 | 0.1801 | 0.1191 | 0.1507 | 0.1264 |
| 8192,3,64 | 0.1177 | 0.07119 | 0.1021 | 0.07696 |
| 8192,3,128 | 0.08575 | 0.04354 | 0.07006 | 0.05057 |
| 8192,3,256 | 0.05213 | 0.03263 | 0.04526 | 0.03157 |
| 8192,4,16 | 0.2458 | 0.1634 | 0.2327 | 0.2001 |
| 8192,4,32 | 0.1755 | 0.1268 | 0.158 | 0.1286 |
| 8192,4,64 | 0.1243 | 0.08322 | 0.1005 | 0.08251 |
| 8192,4,128 | 0.08316 | 0.05443 | 0.07637 | 0.04983 |
| 8192,4,256 | 0.05478 | 0.03888 | 0.04963 | 0.03499 |
| 8192,5,16 | 0.2475 | 0.1915 | 0.2046 | 0.2068 |
| 8192,5,32 | 0.1695 | 0.1139 | 0.177 | 0.1297 |
| 8192,5,64 | 0.116 | 0.07838 | 0.1028 | 0.08718 |
| 8192,5,128 | 0.08529 | 0.05586 | 0.08562 | 0.05641 |
| 8192,5,256 | 0.06015 | 0.04 | 0.04855 | 0.0391 |
| 8192,10,16 | 0.2305 | 0.1811 | 0.2308 | 0.2069 |
| 8192,10,32 | 0.1787 | 0.1383 | 0.1806 | 0.1352 |
| 8192,10,64 | 0.1165 | 0.1005 | 0.1301 | 0.1064 |
| 8192,10,128 | 0.08591 | 0.06522 | 0.0825 | 0.0649 |
| 8192,10,256 | 0.06065 | 0.0462 | 0.05548 | 0.03805 |

Table 7: $\max_{F \in \mathcal{F}} \left| \frac{|F|}{|X|} - \frac{|F \cap A|}{|A|} \right|$ for our algorithms on the grid set system averaged over 10 runs.